# Reduced neutron spectroscopic factors when using potential geometries constrained by Hartree-Fock calculations

Jenny Lee[1], J.A. Tostevin[1,2], B.A. Brown[1], F. Delaunay[1], W.G. Lynch[1], M.J. Saelim[1], M.B. Tsang[1*]

*[1]National Superconducting Cyclotron Laboratory and Department of Physics and Astronomy, Michigan State University, East Lansing, MI 48824*
*[2]Department of Physics, School of Electronics and Physical Sciences, University of Surrey, Guildford, Surrey GU2 7XH, United Kingdom*

## Abstract

We carry out a systematic analysis of angular distribution measurements for selected ground-state to ground-state (d,p) and (p,d) neutron transfer reactions, including the calcium isotopes. We propose a consistent three-body model reaction methodology in which we constrain the transferred-neutron bound state and nucleon-target optical potential geometries using modern Hartree-Fock calculations. Our deduced neutron spectroscopic factors are found to be suppressed by ~30% relative to independent-particle shell-model values, from $^{40}$Ca through $^{49}$Ca. The other nuclei studied, ranging from B to Ti, show similar average suppressions with respect to large-basis shell-model expectations. Our results are consistent with deduced spectroscopic strengths for neutrons and protons from intermediate energy nucleon knockout reactions, and for protons from (e,e′p) reactions, on well-bound nuclei.

The shell model of atomic nuclei was proposed more than half a century ago by Mayer and Jensen [1], for which they shared the 1963 Nobel Prize for Physics. This simple independent-particle picture, that requires each nucleon to occupy a single-particle eigenstate (orbital) in the nuclear mean-field potential, has been immensely robust and successful in describing both systematic and specific features of nuclear structures. Modern shell model calculations also include the effects of residual interactions between these single-nucleon degrees of freedom by the use of microscopic interactions or of effective forces fitted to empirical nuclear masses, charge radii and low-lying excited state spectra, e.g. [2]. Quantifying the ordering, spacing and the resulting distribution of single particle strength among these nucleon states is essential for assessing the evolution of nuclear structures in regions of neutron- and proton-rich nuclei.

These single-particle degrees of freedom are studied quantitatively using direct nuclear reactions, such as single-nucleon transfer, nucleon knockout and elastic break-up reactions. When experiments are well chosen, they can interrogate the dominant strength associated with the state of a single nucleon. More precisely they are sensitive to the single-nucleon overlap functions, the norms of which are the spectroscopic factors (SFs) of the measured transitions, see e.g. [3]. Extensive attempts have been made to deduce nucleon SFs using direct reactions induced by both hadronic and electronic probes. Such analyses are still not fully understood, often revealing model-dependence and raising concerns of the capability to determine SFs in an absolute sense. Transfer reactions continue to be an important technique to populate and to help elucidate the structure of rare nuclei, in particular those produced as lower energy secondary beams. Thus it is important to clarify the techniques for SF determinations using single-nucleon transfer reactions and reduce their uncertainties.

A significant body of (e,e′p) reaction analyses suggest that proton SF values deduced from nuclei near closed shells, including $^{40}$Ca, are suppressed by about 30-40% compared to independent particle model (IPM) expectations [4]. This suppression has been attributed, in different proportions, to short- and medium-range (tensor) nucleon-nucleon correlation effects and to longer-range correlations arising from couplings, by nucleons near the Fermi surface, to collective degrees of freedom. More recently, quantitatively similar suppressions have been required to reconcile measurements of both single-proton and single-neutron knockout reactions from a range of nuclei, both stable [5] and unstable [5,6], with reaction theory predictions.

Historically, conventional distorted waves Born approximation (DWBA) transfer reaction analyses have shown little need for such a systematic suppression of single-particle strength [7, 8]. Marked differences in proton SFs extracted from the (e,e′p) and (d,$^3$He) proton transfer reaction analyses were reconciled by a reanalysis of the (d,$^3$He) data [4], requiring significant changes to the (DWBA) (d,$^3$He) calculations used there. These included fine-tuning of the root mean squared (rms) radii of the active proton single particle orbitals to be consistent with the (e,e′p) data analysis. In this paper we assess related effects in (d,p) and (p,d) neutron transfer reactions by constraining the geometry of the nucleon optical interactions with the target and the active (transferred) neutron orbital rms radii using modern Hartree-Fock (HF) calculations [9]. In this way we not only reduce long-standing parameter ambiguities but also introduce into the transfer reaction description the excellent systematic behavior manifested by the HF calculations across extended mass regions [10-12]. This agreement between the systematics of the HF predictions and a range of nuclear size parameters suggests that the theory should also provide a good description of the spatial extent of individual single-particle states.

We thus propose the following consistent three-body analysis of ground-state-to-ground-state neutron transfer reaction data taking HF theoretical input. We calculate the transfer reaction amplitudes using the Johnson-Soper (JS) adiabatic approximation to the neutron, proton and target three-body system [13]. By this means we include the effects of the break-up of the deuteron in the field of the target and of the transfer of the neutron into (or out of) the breakup continuum. Formulating the reactions from this three-body perspective has the enormous advantage that one needs to specify only (the far better understood) nucleon-target optical interactions. Additionally, we avoid the ambiguity in optical potentials obtained from individual best fits to elastic scattering data, by using global nucleon optical potential descriptions that can be applied consistently at all the required incident energies and for all target nuclei. These optical potentials are derived from the nuclear matter effective nucleon-nucleon interaction of Jeukenne, Lejeune and Mahaux (JLM) [14].

The resulting energy- and density-dependent effective interactions are folded with the target one-body densities, using the mid-point local density prescription [15]. These required densities are taken from Hartree-Fock (HF) calculations based on a Skyrme parameterization that offers quantitative agreement with experimental nuclear size parameters. Specifically, we use the recent SkX parameter set [9], determined from a large set of data on spherical nuclei, including nuclei far from stability. The parameter set accounts for the binding energy differences of mirror nuclei [10], interaction cross sections [11], and nuclear charge distributions [12]. The computed neutron and proton HF densities were used individually in evaluating the isovector contribution to the JLM optical potentials. We adopt the conventional scale factors for the computed real and imaginary parts of the JLM nucleon optical potentials, $\lambda_V$=1.0 and $\lambda_W$ = 0.8, found to be consistent with an analysis of data on several systems [15].

The remaining critical ingredients to the reaction calculations are the geometries of the potentials used to generate the neutron overlap functions. For comparison purposes, in the following we will use two different local potential prescriptions. Potential Set I is just a conventional Woods-Saxon potential of *fixed* radius and diffuseness parameters, $r_0$=1.25 fm and $a_0$=0.65 fm. This geometry has been used extensively throughout the literature on single-nucleon transfer. The aforementioned agreement of the systematics of the Skyrme SkX HF predictions with nuclear size parameters suggests that this theory will also give a good description of individual single-particle states. So, for potential Set II we continue to fix the diffuseness parameter at $a_0$=0.65 fm, to which the calculations are rather insensitive, but then adjust the radius parameter, $r_0$, for each reaction, to reproduce the root mean squared (rms) radius of the relevant transferred neutron orbital, as given by the HF calculation. More precisely, $r_0$ is adjusted so that the mean squared radius of the transferred neutron orbital is $\langle r^2\rangle = [A/(A-1)]\, \langle r^2\rangle_{HF}$, where $\langle r^2\rangle_{HF}$ is the HF calculation value. This adjustment is carried out using the HF separation energy. This small mass correction factor corrects the fixed potential center assumption used in the HF approach.

This theoretical guidance on the spatial extension of the neutron bound-state wave function is critical in our analysis. The sensitivity of the calculated cross-sections, and hence the deduced SFs, to this wave function is primarily to its rms radius. As typical we consider the $^{40}$Ca(d,p) reaction at 20 MeV. Based on calculations using a range of binding potentials, with $1.2 \le r_0 \le 1.3$ fm, $0.6 \le a_0 \le 0.7$ fm and $0 \le V_{so} \le 6$ MeV, that generate $1f_{7/2}$ states with rms radii in the range 3.8 to 4.1 fm, the changes in the computed SFs are reproduced to better than 1.5% by the following finite difference formula,

$$\frac{\delta(\mathrm{SF})}{\mathrm{SF}} = -7.658\,\frac{\delta(\mathrm{rms})}{\mathrm{rms}} - 0.717\,\frac{\delta(a_0)}{a_0}\ . \qquad (1)$$

There is negligible (explicit) dependence on $r_0$ and $V_{so}$ beyond their effects on the rms radius. Thus, changes of 7.7% in $a_0$ [$\delta(a_0)$~0.05 fm] and 2.5% in the rms radius [$\delta$(rms) ~ 0.1 fm] translate into $\delta$(SF) of 5% and 19%, respectively, for this transition.

For the $^{40\text{-}45,47\text{-}49}$Ca isotopes the deduced $r_0$ values decrease monotonically with increasing A, and are 1.343, 1.282, 1.276, 1.270, 1.265, 1.259, 1.250, 1.245, and 1.134 fm, respectively. The most significant and rapid changes are at the start of a new sub-shell, $1f_{7/2}$ (N=21) and $2p_{3/2}$ (N=29). Using these values and the JLM optical potentials, we place constraints on the spatial extension of the (structural) overlap function and on the (dynamical) nucleon optical potentials. In doing so, we expect to determine more consistently and precisely that part of the neutron overlap function that is sampled within the transfer reaction transition amplitude. We have also constrained all significant bound state and optical potential parameters theoretically.

These define the key inputs to the reaction. For both binding potential choices the depths of the central potential wells are adjusted to reproduce the experimental separation energies to ensure the correct asymptotic form of the overlap functions. A spin-orbit potential of strength 6 MeV, with the same (central) geometry parameters, $r_0$ and $a_0$, was included in potential Set II. All calculations treated finite range effects using the local energy approximation (LEA) [16] with the transfer strength ($D_o^2$=15006.25 MeV$^2$ fm$^3$) and range ($\beta$=0.7457 fm) parameters of the Reid soft-core $^3S_1$-$^3D_1$ neutron-proton interaction [17]. Non-locality corrections, with range parameters of 0.85 fm and 0.54 fm [18], were included in the proton and deuteron channels, respectively. The transfer reaction calculations were carried out using a version of the code TWOFNR [19]. The neutron SF was extracted by fitting the theoretical calculations to the first maximum in the measured angular distributions. The errors on the deduced SF are assigned as discussed in Ref. [20].

Before we examine a broader range of targets, we first focus on available data for the $^{40-45,47-49}$Ca isotopes. Since $^{40}$Ca is doubly magic, with closed proton and neutron *sd*-shells, the additional neutrons in $^{41-48}$Ca fill the $f_{7/2}$ orbit. These valence neutron wavefunctions in the Ca isotopes are expected to be good single particle orbitals. Indeed, the predicted SF values from large-basis shell-model (LB-SM) calculations, which include configuration mixing, and the independent particle model (IPM), which neglects such effects, are essentially equal in the calcium isotopes. As most of the (e,e′p) SF analyses are compared to the IPM, we also compare our deduced SFs for the calcium isotopes with those of the IPM. For *n* valence nucleons, each of total angular momentum *j*, the IPM predictions are [21]

$$\mathrm{SF(IPM)} = n, \text{ for } n\text{=even}; \qquad \mathrm{SF(IPM)} = 1-\frac{n-1}{2j+1}, \text{ for } n\text{=odd}. \qquad (2)$$

For $^{40-49}$Ca, these SFs are 4, 1, 2, ¾, 4, ½, 6, ¼, 8 and 1, respectively; the odd/even effects arise from pairing. Figure 1 now shows the ratios of the extracted SFs to these SF(IPM) as a function of mass number A.

The solid stars in Figure 1 represent the ratios SF(HF)/SF(IPM) for the calcium isotopes. These SF(HF) are the new results of our constrained analysis, using the HF-inspired neutron binding potential geometries, Set II, and the JLM nucleon optical potentials obtained using the HF densities of the targets. We observe an overall reduction in the SF(HF) compared to the IPM values of about 25-30%. The additional data point for A=40 (open circle) is the proton SF value, 0.645(50), as deduced from the (e,e′p) analysis of Ref. [4]. Within the assigned experimental uncertainties, the neutron SF(HF) and the proton SF(e,e′p) for $^{40}$Ca agree.

For comparison, we also extract SF values from a conventional adiabatic three-body model reaction analysis, now using the Chapel Hill (CH89) global phenomenological nucleon-target interactions [22] and the standard binding potential geometry, (Set I). The open stars in Fig. 1 show the corresponding SF ratios, SF(conv)/SF(IPM). The ratios for these latter calculations are close to unity within experimental uncertainties although three odd-A isotopes, $^{43}$Ca, $^{45}$Ca, and $^{49}$Ca are somewhat suppressed. The suppression for $^{49}$Ca may be traced to a sharp increase in the rms radius of the $2p_{3/2}$ orbital in $^{49}$Ca (4.59 fm), compared to that of the $1f_{7/2}$ orbit (3.99 fm) in neighboring $^{48}$Ca, when using the standard geometry. However this explanation cannot address the reduction in the SFs for the $^{43}$Ca and $^{45}$Ca nuclei.

The deduced SF(HF) are consistently reduced compared to the SF(conv). The reduction of the SF(HF) values comes from both the changes of optical potential and the use of more realistic (larger) neutron-bound state wave-functions. On average, 15% of the reduction arises from the use of the JLM potential instead of the CH89 global potential. Similar effects were observed in Ref. [23]. The rms radii of the neutron bound state wave functions from Set II, based on the Skyrme SkX HF predictions, are also, on average, about 2% larger than the rms radii from Set I, the conventional Woods-Saxon potential of fixed radius and diffuseness parameters. This results in further reduction of the SF values by about 15%, as was discussed earlier in connection with Eq. (1). The observed suppression is thus a manifestation of both effects. As was stated earlier, we believe that these changes, constrained by the same (HF) theoretical systematics, will better determine the all-important overlap of the distorted waves and bound state wave functions at the nuclear surface.

The Ca isotope SF(conv) values are a subset of a recent large-scale survey of 80 nuclei, studied via the (p,d) and (d,p) transfer reactions [8,20]. In the survey, it was shown that within experimental and theoretical uncertainties, most extracted SF(conv) values, like those for the Ca isotopes, agreed with the predictions of the large-basis shell-model (LB-SM). To examine whether or not the reductions in the deduced SF(HF) are limited to the calcium isotopes, we have applied the same HF-constrained analysis to a selection of the 80 nuclei studied in Ref. [8, 20]. As the HF is less appropriate for the description of single-particle configurations of very light systems, we limit the analysis to A>11. Additionally, beyond the calcium isotopes, the IPM does not take proper account of configuration mixing effects, so we now compare the extracted SF(HF) to large-basis shell model SF(LB-SM) predictions which are calculated with the code OXBASH [24]. These ratios are listed in Table 1 and shown in Figure 2, as a function of the difference between the neutron and proton separation energies in the nuclei concerned, $\Delta S$ ($\Delta S=S_n-S_p$ for neutron SF and $\Delta S=S_p-S_n$ for proton SF). Here, $\Delta S$ is the difference of the neutron and proton Fermi surfaces. For clarity, only those points with an overall uncertainty of less than 25% are included. Data with uncertainties much larger than 20% (the random error assigned to each measurement) have quality control problems in the evaluation. In such case, there is either (a) no second measurement to corroborate the validity of a data set or (b) the standard deviations of the measurements used to extract the SF values are larger than 25%. For the (statistically most significant) cases presented, we note once again an overall SF(HF) reduction of order 30% compared to the SF(conv) of Ref. [8, 20], but with significant residual fluctuations between the values for different nuclei. For the nuclei investigated here, there is no evident dependence of the observed reduction factors on $\Delta S_p$. Limiting the observations to the calcium isotopes (the solid stars in Fig. 2), which span neutron-proton separation energy differences from –11.3 MeV ($^{49}$Ca) to 7.3 MeV ($^{40}$Ca), one draws the same conclusion.

The open circles in Figure 2 are the corresponding ratios of the proton ground state SF for $^{7}$Li, $^{12}$C, $^{16}$O, $^{30}$Si, $^{31}$P, $^{40}$Ca, $^{48}$Ca, $^{51}$V, $^{90}$Zr, and $^{208}$Pb (as listed in Table 1), studied with the (e,e′p) reaction [4, 25]. Similarly, the solid triangles show the ratios of the deduced SF to SF(LB-SM) values from both exclusive and inclusive studies of intermediate energy nucleon knockout reactions. Neutron (proton) knockout values are shown as inverted (upright) triangles (and listed in Table 1). These include, at the extremes of the $|S_n$-$S_p|$ scale, $^{15}$C [26], $^{22}$O [27], $^{34}$Ar [28], and $^{46}$Ar [29], while the values for $^{8}$B, $^{9}$C [30], $^{32}$Ar [27], $^{12}$C, $^{16}$O [5], and $^{57}$Ni [31] overlap the ΔS values of both the transfer and the (e,e′p) analyses. In the case of the inclusive knockout reaction analysis of Ref. [5] and [31], effective neutron and proton removal energies were used, obtained by weighting the physical separation energies to each final state by the corresponding cross-sections. The suppression with respect to the SF(LB-SM) is similar from the three different reactions within the ΔS region in which they overlap. A dependence of the suppression on ΔS is indicated by the nucleon-knockout data that extend the data set into regions of significant neutron and proton asymmetry.

In summary, we have presented a consistent analysis of ground-state-to-ground-state single neutron transfer reaction data using a three-body reaction model that constrains the nucleon bound state and nucleon-target optical potential geometries using modern Hartree-Fock calculations. The methodology removes significant, long-standing potential parameter ambiguities from the reaction analysis through the use of theoretical densities and single-particle orbital rms radii. In so doing, we believe that we have defined more precisely that fraction of the neutron overlap function that is sampled in the transfer reaction amplitudes. The deduced spectroscopic factors SF(HF) for the calcium isotopes, and more generally for other systems, show a reduction of ~30% compared to both shell-model values and the SF(conv) deduced using a global optical potential and a conventional, fixed bound state potential geometry. There is no evidence that the reduction factor is correlated with the nucleon separation energy difference, ΔS, over the range of values available to this (ground state) transfer analysis. This observation, and the observed suppression factor of about 70%, are consistent with deduced spectroscopic strengths for neutrons and protons from intermediate energy nucleon knockout reactions, and for protons from (e,e′p) reactions on well-bound nuclei.

This work was supported by the National Science Foundation under Grants No. PHY-01-10253, No. PHY-02-44453 and by the UK Engineering and Physical Sciences Research Council (EPSRC) under Grant No. EP/D003628.

**Table I**: List of isotopes plotted in Figure 2. $J^{\pi}$ is the angular momentum and parity of the transferred nucleon. For the (p,d), (d,p) [8, 20] and (e,e′p) [4, 25] reactions, only ground state SFs are extracted. The theoretical SF values are obtained from the large-basis shell-model code, OXBASH [5,20,24]. For the neutron and proton knockout reactions [26-31], the deduced quantities are the cross section reduction factors, $R_s$, which are equivalent to SF(expt)/SF(LB-SM).

| (p,d) (p,d) | $j^{\pi}$ | Sn-Sp | SF (Expt) | SF (LB-SM) | SF/SF(LB-SM) |
|---|---|---|---|---|---|
| $^{12}$B | 1/2$^-$ | -10.72 | 0.40 ± 0.06 | 0.83 | 0.48 ± 0.07 |
| $^{12}$C | 3/2$^-$ | 2.75 | 2.16 ± 0.25 | 2.85 | 0.76 ± 0.09 |
| $^{13}$C | 1/2$^-$ | -12.58 | 0.54 ± 0.07 | 0.61 | 0.88 ± 0.12 |
| $^{14}$C | 1/2$^-$ | -12.65 | $1.07 \pm 0.22$ | 1.73 | $0.62 \pm 0.12$ |
| $^{14}$N | 1/2$^-$ | 3.00 | $0.48 \pm 0.08$ | 0.69 | $0.69 \pm 0.11$ |
| $^{15}$N | 1/2$^-$ | 0.62 | $0.93 \pm 0.15$ | 1.46 | $0.64 \pm 0.10$ |
| $^{16}$O | 1/2$^-$ | 3.53 | $1.48 \pm 0.16$ | 2.00 | $0.74 \pm 0.08$ |
| $^{17}$O | 5/2$^+$ | -9.64 | $0.75 \pm 0.10$ | 1.00 | $0.75 \pm 0.10$ |
| $^{18}$O | 5/2$^+$ | -7.90 | $1.46 \pm 0.17$ | 1.58 | $0.92 \pm 0.11$ |
| $^{19}$O | 5/2$^+$ | -13.12 | $0.35 \pm 0.05$ | 0.69 | $0.51 \pm 0.07$ |
| $^{25}$Mg | 5/2$^+$ | -4.73 | $0.21 \pm 0.02$ | 0.34 | $0.61 \pm 0.07$ |
| $^{26}$Mg | 5/2$^+$ | -3.06 | $1.83 \pm 0.38$ | 2.51 | $0.73 \pm 0.15$ |
| $^{27}$Al | 5/2$^+$ | 4.79 | $0.93 \pm 0.13$ | 1.10 | $0.84 \pm 0.12$ |
| $^{28}$Al | 1/2$^+$ | -1.82 | $0.57 \pm 0.08$ | 0.60 | $0.95 \pm 0.14$ |
| $^{30}$Si | 1/2$^+$ | -2.90 | $0.55 \pm 0.07$ | 0.82 | $0.67 \pm 0.08$ |
| $^{31}$Si | 3/2$^+$ | -7.78 | $0.42 \pm 0.07$ | 0.58 | $0.72 \pm 0.11$ |
| $^{32}$P | 1/2$^+$ | -0.71 | $0.39 \pm 0.07$ | 0.60 | $0.65 \pm 0.11$ |
| $^{34}$S | 3/2$^+$ | 0.54 | $1.11 \pm 0.27$ | 1.83 | $0.61 \pm 0.15$ |
| $^{37}$Ar | 3/2$^+$ | 0.08 | $0.27 \pm 0.04$ | 0.36 | $0.74 \pm 0.10$ |
| $^{40}$Ca | 3/2$^+$ | 7.31 | $3.20 \pm 0.46$ | 4.00 | $0.80 \pm 0.11$ |
| $^{41}$Ca | 7/2$^-$ | -0.53 | $0.73 \pm 0.04$ | 1.00 | $0.73 \pm 0.04$ |
| $^{42}$Ca | 7/2$^-$ | 1.20 | $1.31 \pm 0.12$ | 1.81 | $0.72 \pm 0.06$ |
| $^{43}$Ca | 7/2$^-$ | -2.75 | $0.44 \pm 0.05$ | 0.75 | $0.59 \pm 0.07$ |
| $^{45}$Ca | 7/2$^-$ | -4.88 | $0.26 \pm 0.04$ | 0.50 | $0.52 \pm 0.07$ |
| $^{47}$Ca | 7/2$^-$ | -6.93 | $0.19 \pm 0.03$ | 0.26 | $0.74 \pm 0.10$ |
| $^{48}$Ca | 7/2$^-$ | -5.86 | $5.41 \pm 1.05$ | 7.38 | $0.73 \pm 0.14$ |
| $^{49}$Ca | 3/2$^-$ | -11.30 | $0.74 \pm 0.08$ | 0.92 | $0.81 \pm 0.08$ |
| $^{46}$Ti | 7/2$^-$ | 2.85 | $1.61 \pm 0.23$ | 2.58 | $0.62 \pm 0.09$ |
| **(e,e'p)** | **$j^{\pi}$** | **Sp-Sn** | **SF (Expt)** | **SF (LB-SM)** | **SF/SF(LB-SM)** |
| $^{7}$Li | 3/2$^-$ | 2.73 | $0.42 \pm 0.04$ | 0.67 | $0.63 \pm 0.06$ |
| $^{12}$C | 3/2$^-$ | -2.75 | $1.72 \pm 0.11$ | 2.85 | $0.60 \pm 0.04$ |
| $^{16}$O | 1/2$^-$ | -3.53 | $1.27 \pm 0.13$ | 2.00 | $0.64 \pm 0.07$ |
| $^{30}$Si | 5/2$^+$ | 2.90 | $2.21 \pm 0.20$ | 3.80 | $0.58 \pm 0.05$ |
| $^{31}$P | 0$^+$ | -5.01 | $0.40 \pm 0.03$ | 0.58 | $0.68 \pm 0.04$ |
| $^{40}$Ca | 3/2$^+$ | -7.3 | $2.58 \pm 0.19$ | 4.00 | $0.65 \pm 0.05$ |
| $^{48}$Ca | 1/2$^+$ | 5.86 | $1.07 \pm 0.07$ | 1.98 | $0.54 \pm 0.04$ |
| $^{51}$V | 7/2$^-$ | -2.99 | $0.37 \pm 0.03$ | 0.75 | $0.49 \pm 0.04$ |
| $^{90}$Zr | 1/2$^-$ | -3.62 | $0.72 \pm 0.07$ | 1.28 | $0.56 \pm 0.05$ |
| $^{208}$Pb | 1/2$^+$ | 0.63 | $0.98 \pm 0.09$ | 2.00 | $0.49 \pm 0.05$ |
| **n-knockout** | **$j^{\pi}$** | **Sn-Sp** | **SF (Expt)** | **SF (LB-SM)** | **$R_s$** |
| $^{12}$C | incl | 3.07 | | - | $0.49 \pm 0.02$ |
| $^{15}$C | 1/2$^+$ | -19.86 | | 0.98 | $0.96 \pm 0.04$ |
| $^{16}$O | incl | 7.64 | | - | $0.56 \pm 0.03$ |
| $^{22}$O | 5/2$^+$ | -16.39 | | 5.22 | $0.70 \pm 0.06$ |
| $^{32}$Ar | 5/2$^+$ | 19.20 | | 4.12 | $0.25 \pm 0.03$ |
| $^{34}$Ar | incl | 13.94 | | - | $0.41 \pm 0.07$ |
| $^{46}$Ar | 7/2$^-$ | -10.63 | | 5.41 | $0.85 \pm 0.12$ |

| $^{57}$Ni | incl | 5.97 | | - | 0.51 ± 0.02 |
|---|---|---|---|---|---|
| **p-knockout** | **$j^{\pi}$** | **Sp-Sn** | **SF (Expt)** | **SF (LB-SM)** | **$R_s$** |
| $^{8}$B | incl | -12.82 | | - | 0.86 ± 0.07 |
| $^{9}$C | $3/2^{-}$ | -12.96 | | 0.94 | 0.82 ± 0.06 |
| $^{12}$C | incl | -2.43 | | - | 0.53 ± 0.02 |
| $^{16}$O | incl | 0.68 | | - | 0.68 ±0.04 |

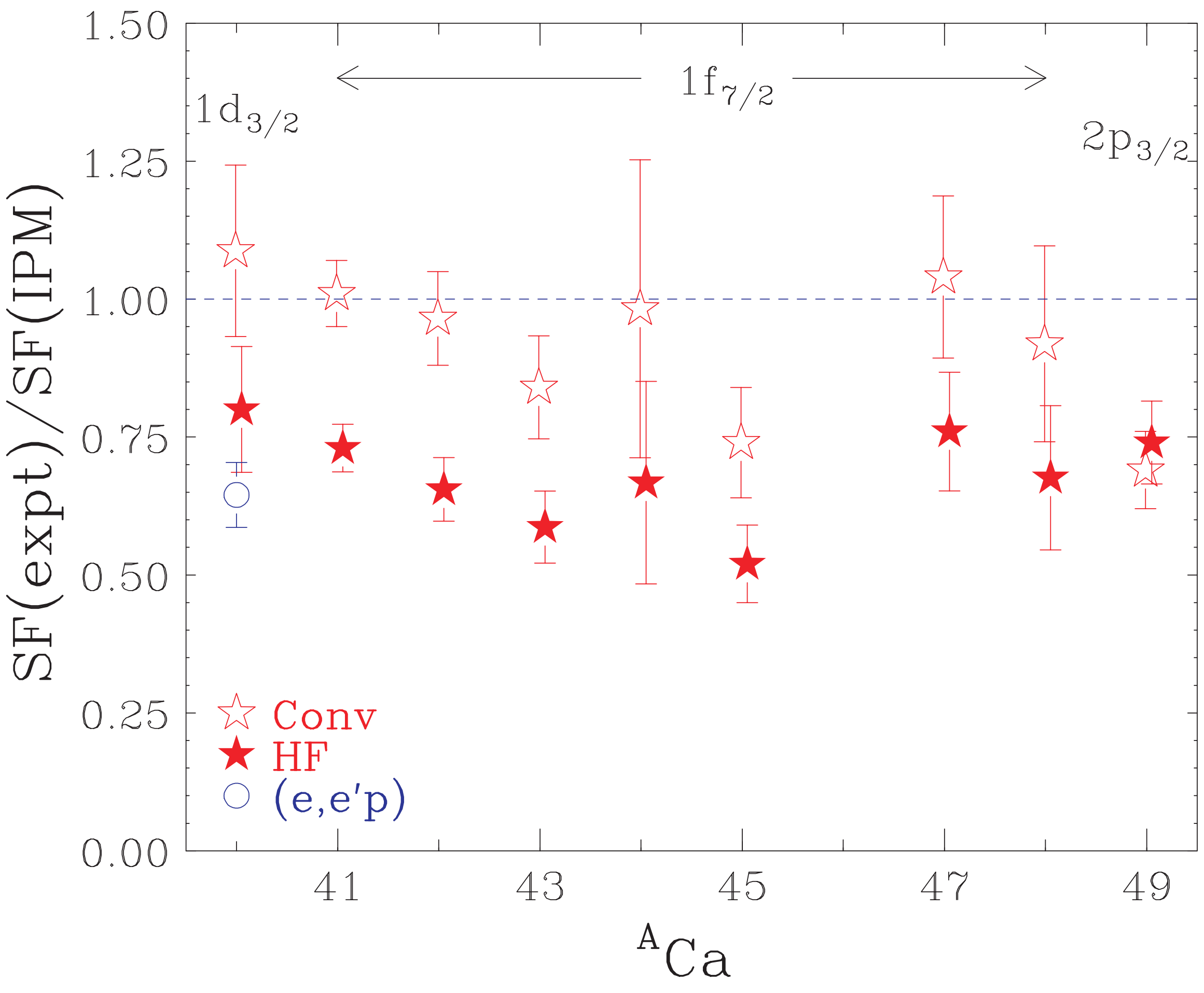

**Figure 1**: (Color online) Ratios of the experimentally deduced spectroscopic factors to those of the independent particle shell-model SF(IPM) for the calcium isotopic chain. The open symbols, from SF(conv), result from the use of conventional, three-body adiabatic model calculations using the Chapel Hill global nucleon optical potentials and a fixed neutron bound-state geometry (Set I) [5]. The solid symbols, from SF(HF), are the results of constrained three-body model calculations, where both the nucleon optical potentials (the JLM microscopic optical model) and the neutron bound state potential geometry (Set II) are determined by the Skyrme (SkX) Hartree-Fock calculations of Ref. [9].

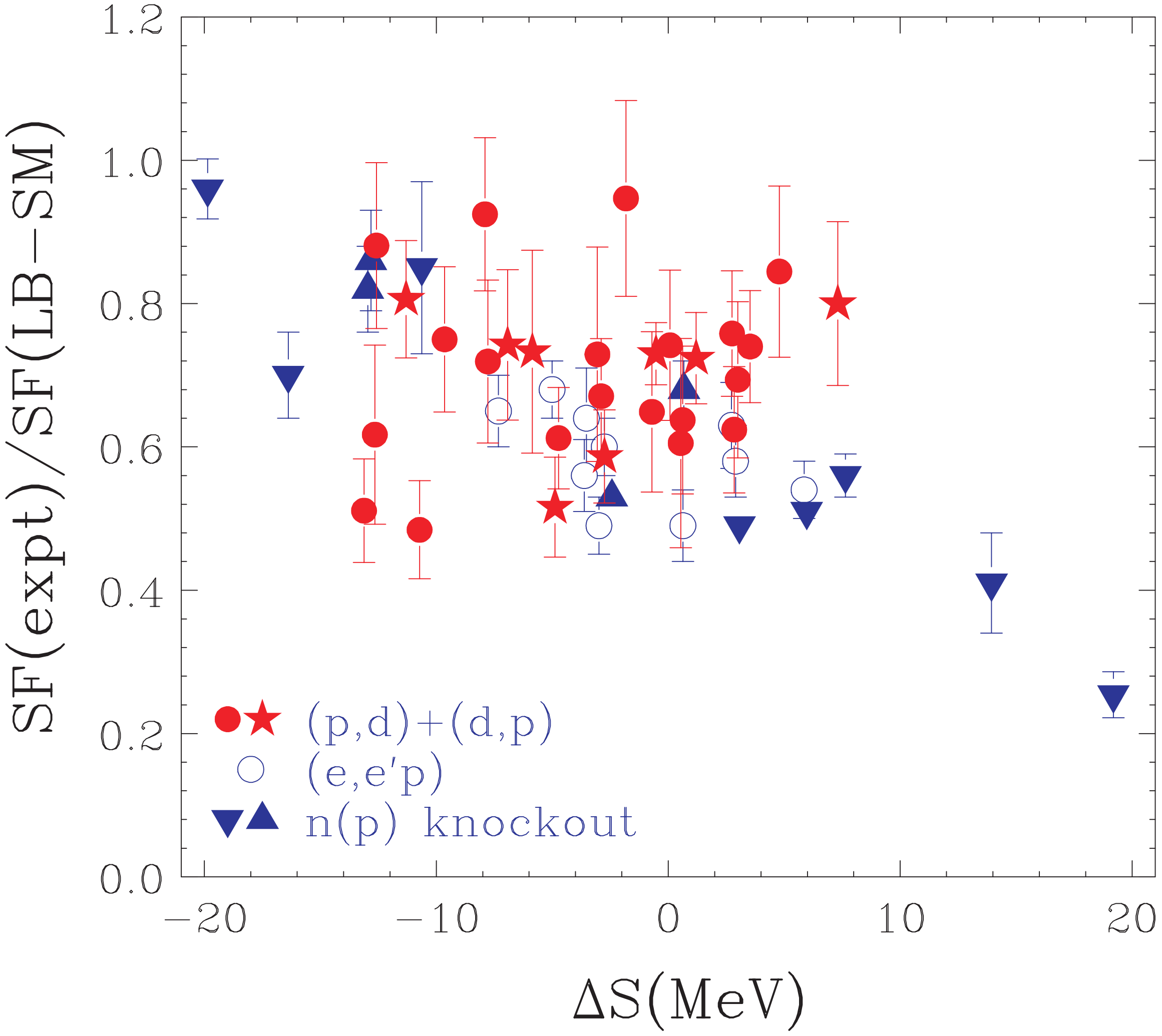

**Figure 2**: (Color online) Ratios of experimentally deduced SF to those of the large-basis shell-model calculations SF(LB-SM) for nuclei with A=12-49 as a function of the difference of neutron and proton separation energies, ΔS (see text). The solid circles and stars are the present results from transfer reactions. The solid stars represent the Ca isotopes, as in Figure 1. The open circles are ground state proton SF from (e,e′p) analysis and the triangles are the results from proton knockout reactions (with inverted triangles for neutron knockout). The data points are listed in Table I and are referenced in the text.